\documentclass[reqno,showpacs,nofootinbib,12pt]{revtex4}

\usepackage{graphicx}
\usepackage{epsfig}

\usepackage[centertags]{amsmath}
\usepackage{amsfonts,amsmath}
\usepackage{amssymb}
\usepackage{amsthm}
\usepackage{newlfont}
\usepackage{mathrsfs}

 \let\la=\lambda

\newcommand{\opunit}{\text{1}\kern-0.22em\text{l}}

\DeclareMathAlphabet{\mathpzc}{OT1}{pzc}{m}{it}

\newcommand{\id}{\textrm{d}}

\newcommand{\bbZ}{{\mathbb Z}}

\def\bea{\begin{eqnarray}}
\def\eea{\end{eqnarray}}
\def\ba{\begin{array}}
\def\ea{\end{array}}
\def\n{\nonumber}
\def\c{\mathscr}
\def\la{\langle}
\def\ra{\rangle}

\usepackage{changes}

\begin{document}

\title{The frenetic origin of negative differential response}
\author{Pieter Baerts}
\author{Urna Basu}
\author{Christian Maes}
\email{christian.maes@fys.kuleuven.be}
\author{Soghra Safaverdi}
\affiliation{Instituut voor Theoretische Fysica, KU Leuven, Belgium}


\begin{abstract}
The Green-Kubo formula for linear response coefficients gets modified when dealing with  nonequilibrium dynamics.  In particular negative differential conductivities are allowed to exist away from equilibrium.  We give a unifying framework for such negative differential response in terms of the frenetic contribution in the nonequilibrium formula. It corresponds to a negative dependence of the escape rates and reactivities on the driving forces.  Partial caging in state space and reduction of dynamical activity with increased driving cause the current to drop.  These are time-symmetric kinetic effects that are believed to play a major role in the study of nonequilibria. We give various simple examples  treating particle and energy transport, which all follow the same pattern in the dependence of the dynamical activity on the nonequilibrium driving, made visible from recently derived nonequilibrium response theory. 
\end{abstract}

\maketitle

\section{Introduction}
Green--Kubo formul{\ae} \cite{Green1954, Kubo1957,mori,lar}  relate equilibrium fluctuations to conductivities  of an equilibrium system.  They allow to compute the more microscopic dependence of the current on the force, summarized in terms of current-current correlations for the linear response coefficients. Their positivity follows often by inspection, e.g. by rewriting them as Helfand moments generalizing the Sutherland--Einstein relation between mobility and diffusion constant \cite{helf}.   Main examples include the positive conductance expressing Ohm's law, the strain rate for mechanical transport following Newton's law, the thermal conductivity in Fourier's law, etc.  There are also deeper reasons of thermodynamic stability why some of these coefficients must always be positive.   The thermodynamic stability refers in the first place to the positivity of the entropy production.  Within the context of irreversible thermodynamics, the argument runs as follows.  Currents $J_i$ are linearly related to 
forces $F_i$ with Onsager response matrix $L$:
\[
J_i = \sum_{j} L_{ij} F_j 
\]
making the entropy production equal to $\sigma = \sum_i J_i F_i = \sum_{ij} F_i F_j L_{ij}$.  Asking $\sigma\geq 0$ is equivalent to requiring that the Onsager matrix $L$ be positive.

When away from thermodynamic equilibrium, the linear response coefficients (around nonequilibrium) need not give rise to a  positive linear response matrix (even though the entropy production of course remains positive).  And indeed many physical systems with negative differential response have been  observed and investigated.  Most of these studies have however remained with a specific model or type of mechanism for the particular context.  Here we attempt a unifying theory where  negative response is understood from a correlation between the current and the dynamical activity.  That is the frenetic origin to which the title alludes, to be illustrated by a choice of examples in the following sections and which we discuss  in the last section from a more general perspective.  The logic can be summarized as follows.  For perturbations around nonequilibrium the response is no longer given {\it only} via the standard Kubo formula; there is a second {\it frenetic} contribution in the form of a 
correlation  $\langle J \text{d}D\rangle$  
between the time-antisymmetric current $J$ and the excess dynamical activity $\text{d}D$.    The latter refers to a sort of time-symmetric current, meaning the rate of escape from a given state or reactivity.  When the system shows trapping behavior, e.g. by getting stuck in some phase space cages, the dynamical activity is affected.  If the trapping behavior significantly grows by the perturbation, effectively diminishing the escape of the system, then negative differential response will occur\footnote{Negative differential response is distinct from {\it absolute} negative response where the current flows in the opposite direction of the applied field; see for example  \cite{Cleuren,Reimann,spiechowicz}. In the present paper we choose for models that also have an equilibrium version with corresponding linear response for small driving.}. This picture is quite intuitive and has been suggested before e.g. in \cite{Zia} for an example that we will also meet in Appendix \ref{app:dhooks}; it has inspired us to 
suggest a biased 
random walker as a paradigmatic model of transport 
where also the escape rate (strongly) depends on the biasing field.  That model will be detailed in the next section.  
Such heuristics will be accompanied by a more general and precise formula for nonequilibrium response allowing quantitative studies also in cases where exact results are not available, also reviewed in the next Section \ref{sec:response_formula}.\\

For the plan of the rest of the paper, we basically deal with two types of models, for particle and for energy transport, respectively in Section \ref{sec:partr} and in Section \ref{sec:thertr}\,\footnote{For momentum transport currents are time-symmetric and they require a separate analysis; see also the first remark of Section \ref{sec:rema}.}.  For particle transport we study the biased motion of particles in a medium with obstacles.  A first example is a colloidal particle immersed in an equilibrium fluid, driven through a narrow tube with hooks, i.e., vertical and horizontal spikes partially blocking free streaming.  A second example is a Lorentz lattice gas with driven random walkers on a two-dimensional lattice with random obstacles.  Both examples can be effectively mapped on our paradigmatic model of a one-dimensional biased random walker with field-dependent escape rates.\\
Section \ref{sec:thertr} provides a discrete model of heat conduction and gives a mechanism for negative differential heat conductivity which is again based on trapping.  
Also kinetic 
factors in energy transport are affected by the installed temperature difference.  If, at higher temperature difference, these kinetic factors slow down the transport an opposite tendency to reduce the energy current arises. \\
The last sections take up a more general perspective.  We add various remarks and we attempt a general heuristics in which the frenetic contribution in nonequilibrium is related to a surface effect in abstract phase space, to be compared with volume effects (i.e., entropic forces) in the relaxation to equilibrium.

\section{Modified Green-Kubo formula}

The aim of linear response theory is to predict the change in the expected value of an observable $O$  upon some external stimulus.  The present set-up is to imagine a change $h\rightarrow h+\id h$ in an existing field or potential indicated by $h.$

\subsection{Response formula involving the dynamical activity}\label{sec:response_formula}
 Let us consider an open system in contact with one or different equilibrium reservoirs and/or subject to external forces.  We denote by $x$ the state of the open system, e.g. the position of particles in a medium. 
For each trajectory $\omega := (x_s, 0\leq s \leq t)$ of the system over the interval $[0,t]$ we identify two quantities, the entropy flux $S_h(\omega)$ and the dynamical activity $D_h(\omega)$. The way to compute them for a given dynamical ensemble is described in \cite{fdr,update} and we repeat the main steps in Appendix \ref{app:respa}.   The result is that the differential response to the perturbation is given by
\bea
\frac \id{\id h} \la O(\omega)\ra^h = \frac 12 \left \la O(\o)\frac {\id S_h}{\id h}(\omega)  \right\ra^h - \left \la O(\omega)\frac {\id D_h}{\id h}(\omega)\right\ra^h \;\;\;\; \label{eq:dQSD}
\eea
Putting there $O=1$ we get $\frac 12\left \la \frac {\id}{\id h} S_h(\omega) \right\ra = \left \la \frac {\id}{\id h} D_h(\omega) \right\ra,$ from which we rewrite \eqref{eq:dQSD} as,
\bea
\frac \id{\id h} \la O(\omega)\ra^h = \frac 12 \left \la O(\omega);\frac {\id S_h}{\id h}(\omega) \right\ra^h - \left \la O(\omega);\frac {\id D_h}{\id h}(\omega)\right\ra^h ~~~\label{eq:dQSD_cov}
\eea
$\la A;B \ra$ denotes the covariance between the observables $A,B$.  The averages $\langle\cdot\rangle^h$ are over trajectories including possibly the initial conditions and depending on the considered field $h$.  We will often drop the explicit dependence on $h$ in the notation.   Thus, the first term in \eqref{eq:dQSD_cov} signifies the covariance or the connected correlation of the observable $O$ with the linear excess of entropy generated due to the perturbation and the second term arises from the correlation with the change in dynamical activity.\\

Assuming that $h=0$ corresponds to equilibrium (also including an initial averaging over the equilibrium distribution) and that  the observable $O$ is time-antisymmetric then
\[
\left \la O(\omega) ; \frac {\id}{\id h} D_h|_{h=0}(\omega) \right\ra^0 =0
\]
because the dynamical activity $D_h(\omega)$ in \eqref{eq:dQSD_cov} is itself time-symmetric and equilibrium is time-reversal invariant.  Thence,
\bea
\left.\frac \id{\id h} \la O(\omega)\ra^h\right |_{h=0} = \frac 12 \left \la O(\omega) ; \frac {\id}{\id h} S_h|_{h=0}(\omega) \right\ra^0 \label{eq:equilbis}
\eea
  That equilibrium result \eqref{eq:equilbis} is basically the Green-Kubo relation but we do not rewrite it here by e.g. replacing the entropy flux in terms of currents.  We will see it more explicitly in later examples.\\
 The frenetic contribution $\left \la O(\omega) ; \frac {d}{dh} D_h(\omega) \right\ra^h$ involving the dynamical activity $D(\omega)$ is thus the key term which differentiates nonequilibrium response from that around equilibrium. In particular, a large frenetic contribution  can also result in a negative differential response $\frac d{dh} \la O(\omega)\ra^h \leq 0$ in some regime of the parameter $h,$ even in cases where that is strictly forbidden and not possible in equilibrium.\\

To illustrate the use of words, we make more explicit the entropic and frenetic contributions here for systems modeled by Markov jump processes. 
These are specified by transition rates $k(x,y)$ for jumps $x\rightarrow y$ between states $x,y$.  We parameterize them as
\bea
k(x,y) &=& \psi(x,y)\,e^{s(x,y)/2},\cr
\psi(x,y)&=&\psi(y,x)\geq 0, \; s(x,y)= -s(y,x) \label{eq:psi_s}
\eea
all possibly depending on the field or potential $h$.   A trajectory $\o:=(x_s, 0\le s \le t)$ over time interval $[0,t]$  is characterized by discrete jumps at times $s_i$ and by exponentially distributed waiting times $s_{i+1}-s_i.$  Then, for substituting in \eqref{eq:dQSD_cov} --- see Appendix \ref{app:respa},
\bea
S_h(\omega) &=&  \sum_i s(x_{s_i},x_{s_{i+1}}) \cr
D_h(\omega) &=& \int_0^t \id s \, \xi(x_s) - \sum_{i} \log \psi(x_{s_i},x_{s_{i+1}})  \label{eq:ShDh}
\eea
where $\xi(x) = \sum_y k(x,y)$ is the escape rate at state $x$.  The last line gives the expression for the path-dependent dynamical activity.  Note that it is time-symmetric (reversing the time over the trajectory in $[0,t]$ does not affect it) and that it is characterized by reactivities $\psi$ and escape rates $\xi$. It summarizes  those {\it kinetic} factors that become especially important outside equilibrium. In contrast, $S_h$ is time-antisymmetric and corresponds to the {\it thermodynamic} entropy flux over $[0,t]$ whenever the condition of local detailed balance is verified \cite{time,leb,kls,har,der,hal}.  Then indeed $s(x,y)$ is the entropy flux to the environment (per $k_B$) in the transition $x\rightarrow y$.\\
From \eqref{eq:ShDh} we calculate the excess entropy and dynamical activity produced by the perturbation to be used in \eqref{eq:dQSD_cov} to obtain the linear response. 
 In the following sections we apply this formalism to explain the origin of negative differential response of several systems.  What will happen is summarized in the following simple model.

\subsection{Reference example: biased random walk}\label{sec:brw}

We formulate here the paradigmatic example of negative differential response to which all other examples can somehow be reduced.

Consider a one-dimensional nearest neighbor continuous time random walk specified by  rates $p$ and $q$ of jumping to the right, respectively left neighbor. In the parameterization \eqref{eq:psi_s},
\[
\psi(x,x \pm 1) = \sqrt{pq},\quad s(x,x\pm 1) = \pm \log \frac{p}{q},\quad x\in \bbZ
\]
Equilibrium dynamics corresponds to $p=q$.  
We imagine an external field $E\geq 0$ bringing about the bias $p\geq q$ and working in an environment at constant temperature $\beta^{-1}$ so that we get a physical characterization by putting
\[
p+q = g_\beta(E),\quad \log \frac{p}{q} = \beta E
\]
The function $g_\beta(E)$ gives the dependence of the escape rate $\xi(x) = p+q$ on the field $E$.\\
We look at differential conductivity; how does the particle velocity change by an increase in the field.  For this we use  formula \eqref{eq:dQSD_cov} where we now write $h=E$ and with $O$ being the time-integrated current $J$ (net number of steps to the right).  We find the entropy flux and the dynamical activity from \eqref{eq:ShDh}:
\bea
S(\omega) &=& (N_+-N_-) \log \frac pq = \beta E J \cr
D(\omega) &=& (p+q)t -\frac 12 (N_++N_-) \log pq  \cr
      &=& g_\beta(E)t + N\left[\frac {\beta E}2 +\log(1+e^{-\beta E}) -\log g_\beta(E)\right]\n
\eea
We have indicated the number of jumps $N_+$ and $N_-$ to the right and left respectively. The current $J = N_+-N_-$ and  $N=N_++N_-$ is the total number of jumps during the interval $[0,t].$
The change in the current caused by a small increase in the field $E \to E+\id E$ is expressed as a sum of two terms following Eq. \eqref{eq:dQSD_cov},
\bea
{\id \over \id E}\la J\ra &=& \frac 12   \left \la J; \frac \id{\id E}S(\omega)\right \ra - \left \la J; \frac \id{\id E} D(\omega) \right \ra \cr
&=& \frac \beta 2 \la J;J \ra+\left( \frac {g^\prime_\beta(E)}{g_\beta(E)}-\frac \beta 2 {1- e^{-\beta E} \over 1+ e^{-\beta E}}  \right) \la N;J\ra~ \label{eq:dJdE_rw}
\eea
The first term, variance of the current $J$, is the positive definite entropic contribution whereas the second term  involves the covariance of the current with the total number of jumps,  i.e., with the dynamical activity. 

\begin{figure}[t]
 \centering
 \includegraphics[width=16 cm]{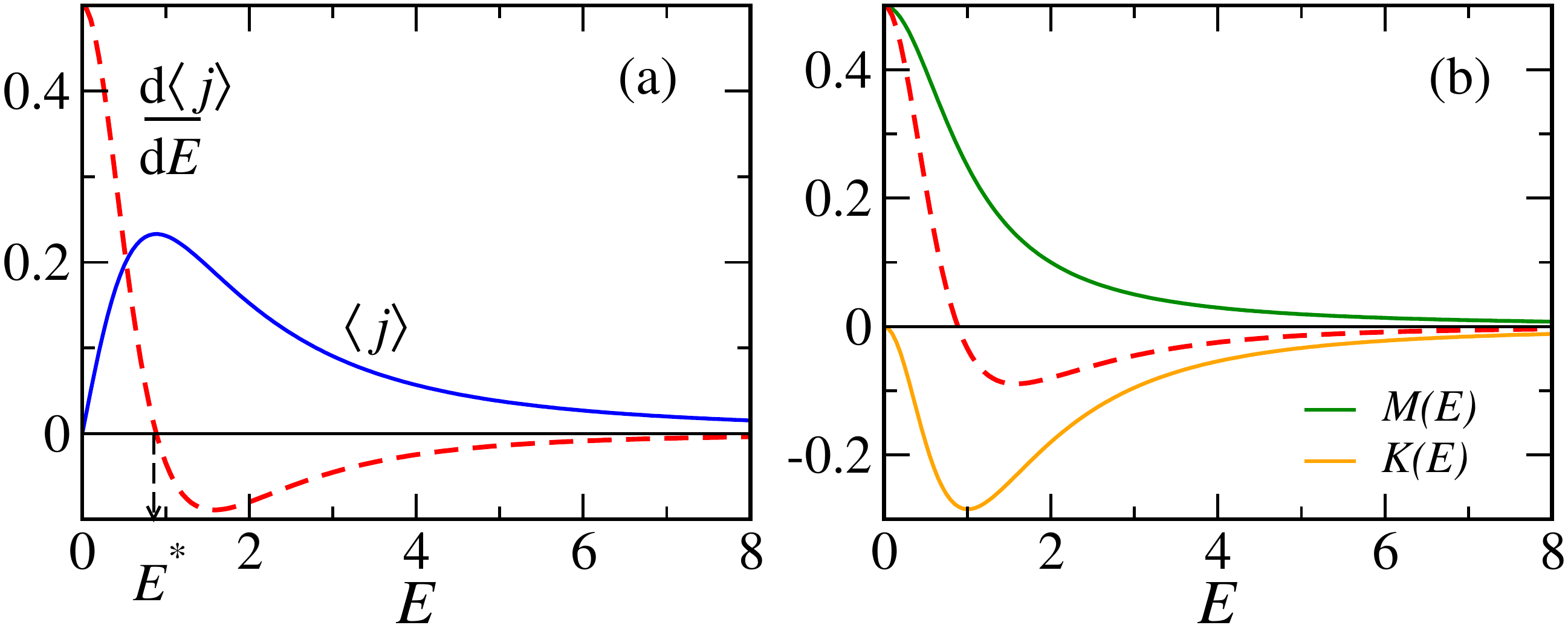}
 \caption{(Color online)(a) The average velocity $\la j\ra$ (solid line) and differential conductivity $\frac {d \la j \ra}{dE}$ (dashed line) as functions of the field $E.$  (b) Plots of the entropic $M(E)$ (upper solid curve) and frenetic contribution $K(E)$ (lower solid curve). The dashed curve is obtained by adding these two and is identical to the one in the left panel.  Here $g_\beta(E)= \frac 1{1+\beta^2 E^2},$ and $\beta=1.$}\label{fig:RW}
\end{figure}

Before we discuss this expression any further and to avoid misunderstanding,  we hasten to add that for the present example all these quantities can be calculated exactly.  There is for example no mystery about the current of the walker; the average current is just
\bea
\frac 1{t}\la J \ra = p-q  =   {1- e^{-\beta E} \over 1+ e^{-\beta E}}\, g_\beta(E) \label{eq:rw_J}
\eea
Clearly the behavior of the current as a function of the external field depends on the nature of the escape rate $g_\beta(E).$
In particular, one can obtain a non-monotonic behavior of the current if $g_\beta(E)$ happens to be a decreasing function of the field $E.$  A decreasing $g_\beta(E)$ signifies  an increase in the degree of trapping of the system. Taking the $E$-derivative of \eqref{eq:rw_J}  obviously verifies formula \eqref{eq:dJdE_rw} as 
we can also calculate separately
\bea
\la J;J \ra &=& \ g_\beta(E)t \cr
\la N;J \ra 
            &=& g_\beta(E)t {1- e^{-\beta E} \over1+ e^{-\beta E}} \label{eq:JJ_NJ_rw}
\eea
The point of the present example is rather that we see so clearly  how the field dependence in the escape rate (trapping mechanism) leads to negative differential response, and how that is exactly picked up by the frenetic contribution in the response formula \eqref{eq:dJdE_rw}.  To be explicit we illustrate all that with the example $g_\beta(E) = {1 \over 1 + (\beta E)^2}.$
Fig. \ref{fig:RW}(a) shows the plot of current and differential conductivity as a function of field strength $E.$ For the sake of convenience here we have used  velocity $i.e.$ current per unit time   $j=J/t$ instead of the time integrated current $J.$ We write the corresponding response as,
\bea
{d \over dE}\la j\ra = M(E) + K(E) 
\eea
where $M(E)$ and $K(E)$ are the entropic and frenetic contributions as calculated from Eq. \eqref{eq:dJdE_rw}. Explicitly,
\bea
M(E) &=& \frac \beta 2 \frac 1{(1+ (\beta E)^2)} \cr
K(E) &=& -\frac \beta 2 \left[ \frac{\beta E}{1 + (\beta E)^2} + {1- e^{-\beta E} \over1+ e^{-\beta E}} \right]{1- e^{-\beta E} \over1+ e^{-\beta E}}\frac{1}{1 + (\beta E)^2} \n
\eea
In this case the frenetic term is negative for all $E>0.$ The variations of the entropic and frenetic contributions with the field strength $E$ are shown separately in Fig. \ref{fig:RW}(b).  The frenetic contribution becomes very negative at around $\beta E=1$ causing the current to drop.\\

Let us look further at more general features of the response formula \eqref{eq:dJdE_rw}.
We are particularly interested in negative differential response. It is clear from equations \eqref{eq:dJdE_rw} and  \eqref{eq:JJ_NJ_rw} that negative ${d \over dE}\la J\ra$ can result only when the coefficient of $\la N;J\ra$ becomes `sufficiently' negative. \\
The critical value $E^*$ at which the conductivity becomes negative depends on the particular choice of $g_\beta(E)$ and temperature $\beta^{-1}.$ Physically we expect as the ambient temperature is increased it would take larger field strength to reach the negative conductivity regime. This can be seen more concretely  when $g_\beta(E)= g(\beta E);$ for that case it is straightforward to  find $E^* \sim \beta^{-1}$ by taking the derivative of Eq. \eqref{eq:rw_J} and equating it to zero.

 Naturally, near equilibrium, the entropic contribution dominates.  We can see it by expanding \eqref{eq:dJdE_rw} around $E=0$:
\[
\frac 1{t}{d \over dE}\la J\ra^E =
\frac \beta{2t} \la J;J \ra^0 + \frac \beta{2} g'_\beta(0) \,E + \ldots  
\]
which is just a small perturbation of the Green-Kubo formula.  The first nonlinearity in the response near equilibrium is thus decided by the derivative of the escape rate $g_\beta(E)$ as function of the field $E$, which can already contribute negatively.
Obviously for large driving field $E$  the frenesy contributes substantially and the response deviates from the Green-Kubo formula. Somewhat surprisingly however, for the special choice $g_\beta(E) = \cosh \beta E/2$ the frenetic term vanishes for all field strengths.  Then, the differential response is always entropic, that is to say it follows the Green-Kubo formula (only the first term in \eqref{eq:dJdE_rw}) even though the system most definitely is driven.\\

 It was already argued by Zia {\it et al.} in \cite{Zia} that a key ingredient to obtain negative response in any dynamical system is the presence of some kind of `trap' in the system. In conformation with this conjecture, we point out that a decreasing $g_\beta(E)$ directly lowers the dynamical activity giving rise to the `trapping' of the system. In the following section we explore a few models which have this feature and show that in each case the dynamics can effectively be mapped to such a biased 1-d random walk with a field dependent escape rate $g_\beta(E).$

\section{Particle transport}\label{sec:partr}

One of the simplest nonequilibrium set-ups is to consider independent particles driven by some external force. The environment is assumed to be in thermal equilibrium at some temperature $\beta^{-1}.$  If the velocity of the particle (or the mass current) decreases when the forcing is increased we speak of a negative differential mobility. For small values of forcing the velocity increases as predicted by the equilibrium linear response relations, but there are simple toy-examples of far from equilibrium systems where  a negative differential mobility is indeed found \cite{Zia, Barma}.

In this section we consider two model systems where a driven particle system shows negative differential conductivity. In each case we show, using numerical simulations, that the negativity of the response originates from the correlation of the current with the  change in dynamical activity of the system.

\subsection{Diffusion of colloids in a narrow tube with hooks}\label{sec:hooks}

Our first example is the motion of a driven Brownian particle through a narrow channel \cite{zwan}. Transport properties of narrow corrugated channels with different shape and geometries have also been investigated in recent years \cite{ghosh2012brownian,ghosh2012driven}. In the following  the channel is  compartmentalized in a specific way so as to facilitate local trapping.  A discrete version, after \cite{Zia}, is presented in Appendix \ref{app:dhooks}.

\begin{figure}[t]
\centering
\includegraphics[width=10 cm]{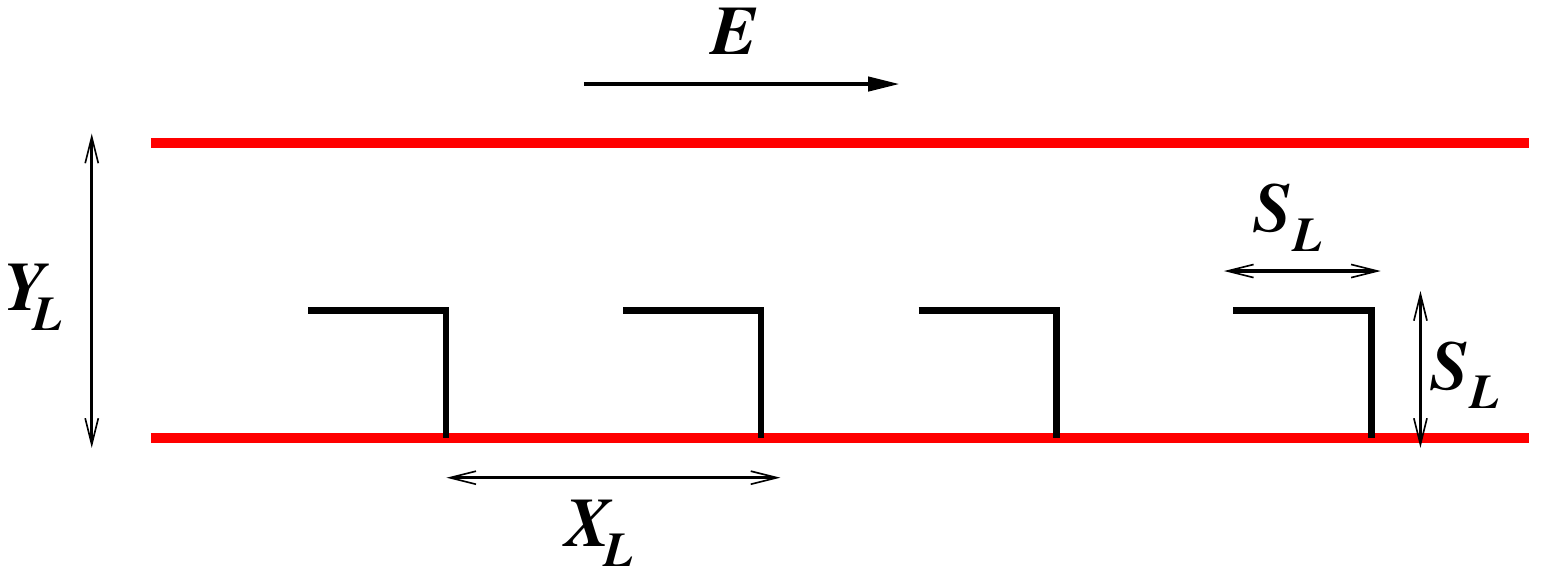}
 \caption{(Color online) Schematic diagram of a narrow tube with hooks attached to it. Each unit cell of the tube has a dimension $X_L \times Y_L.$ The hooks have a linear dimension $S_L.$ The external force $E$ acts along the length of the tube.}
 \label{fig:tube}
 \end{figure}

A point particle of unit mass moves in a fluid contained in a two-dimensional narrow tube of width $Y_L$ with hard, impenetrable and perfectly reflecting walls. The tube is divided in cells by attaching hooks to the lower surface of the tube at regular intervals $X_L.$ The hooks have a linear size $S_L;$ this geometry is illustrated in Fig.~\ref{fig:tube}. The particle is driven by a constant force along the length of the tube. The hooks are expected to provide the trapping mechanism necessary for the negative differential response in the velocity.

The state of the particle at any time $s$ is specified by its position $(x_s,y_s)$ and velocity $(v_{x_s},v_{y_s});$ the surrounding fluid acts as a thermal bath with temperature $T=\beta^{-1}.$ The free dynamics of the  particle is therefore governed by the Langevin equations,
\begin{eqnarray}
\dot x &=& v_x ; \quad\dot {v}_x = -\gamma v_x + \sqrt{\frac{2\gamma}{\beta}} \xi_x + E \cr
\dot y &=& v_y ; \quad\dot {v}_y = -\gamma v_y + \sqrt{\frac{2\gamma}{\beta}} \xi_y
\label{norvin}
\end{eqnarray}
The noises  $\xi_x$ and $\xi_y$ 
are taken to be uncorrelated white noise with zero mean. There is no forcing along the width of the tube.
The constant force $E$ along the length drives the particle to a nonequilibrium condition. We are interested in the response of the velocity of the particle as this force is increased by a small amount. This response is quantified by the differential mobility,
\bea
\mu(E) = \lim_{t\to \infty} \frac{\id}{\id E}\la v_x(t) \ra^{E}
\eea
Another quantity of interest is the diffusion constant, which measures the fluctuation in the position of the particle,
\bea
\mbox{D}_\text{dif}(E) =  \lim_{t\to \infty} \frac 1{2t} \big[\langle (x_t-x_0)^2 \rangle - \langle x_t - x_0 \rangle^2 \big]
\eea

In equilibrium, when there is no forcing, the diffusion constant $D_\text{dif}$ (not to be confused with the dynamical activity $D(\omega)$) and the mobility $\mu$ are related by the Sutherland--Einstein equation $\mu(0) = \beta D_\text{dif}(0)$. In presence of external driving force this relation is no longer valid; mobility and diffusion are not proportional to each other in nonequilibrium situations; see \cite{prs,soghra}.

\begin{figure}[t]
\centering
\includegraphics[width=16 cm]{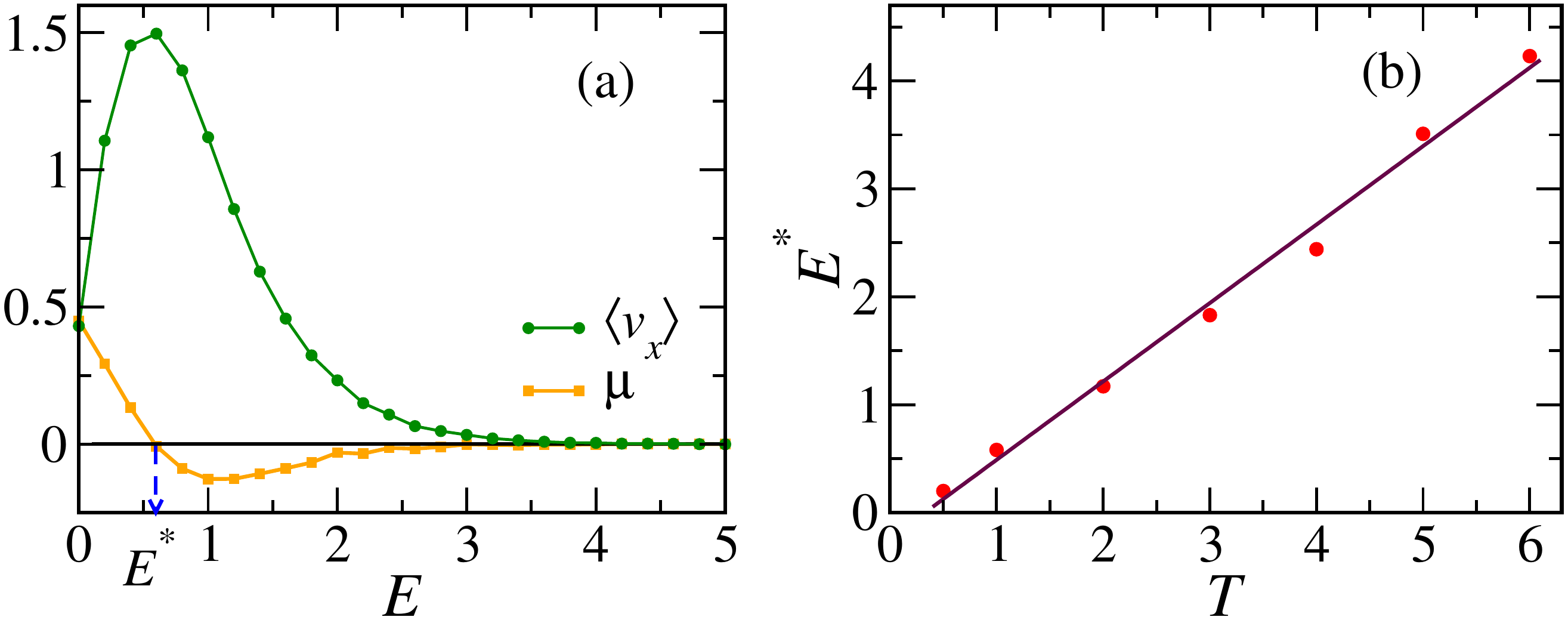}
 \caption{(Color online) (a) The  average current in $x$-direction $\la v_x \ra$ (shown as dark green circles) and mobility (light orange squares) 
in a narrow tube with hooks. (b) The temperature dependence of the critical field $E^*$ after which negative response sets in. For both plots $X_L=Y_L=5, S_L= 2.5,\beta =1$ and $\gamma=1$.}
\label{fig:tube_vel}
\end{figure}

  We use numerical simulations to study the response of this system; Fig. \ref{fig:tube_vel}(a) shows the dependence of $\la v_x\ra$ on the external force $E.$ As $E$ becomes larger the mobility decreases and becomes negative after a certain value $E^*$ which increases linearly with  temperature (see Fig. \ref{fig:tube_vel}(b)). The differential mobility eventually reaches a minimum, increases again and saturates to zero for very large forces. The diffusion constant (not shown) increases initially for small forces and reaches a maximum around the same value where the mobility is minimal!

Physically the negative differential mobility indicates that the particle becomes more trapped in the `cages' as the external force is increased. That is the picture of the biased random walk in Section \ref{sec:brw}.  We can indeed effectively describe it that way, as  illustrated in the next section.  We also checked  that if only vertical obstacles (spikes) are present, then there is no negative mobility; spikes only are not sufficient to trap the particles.  In particular, motion in a channel with hooks as in Fig.~\ref{fig:tube} but with reversed field $E$ will not show a negative differential conductivity.

\begin{figure}[t]
\centering
\includegraphics[width=16 cm]{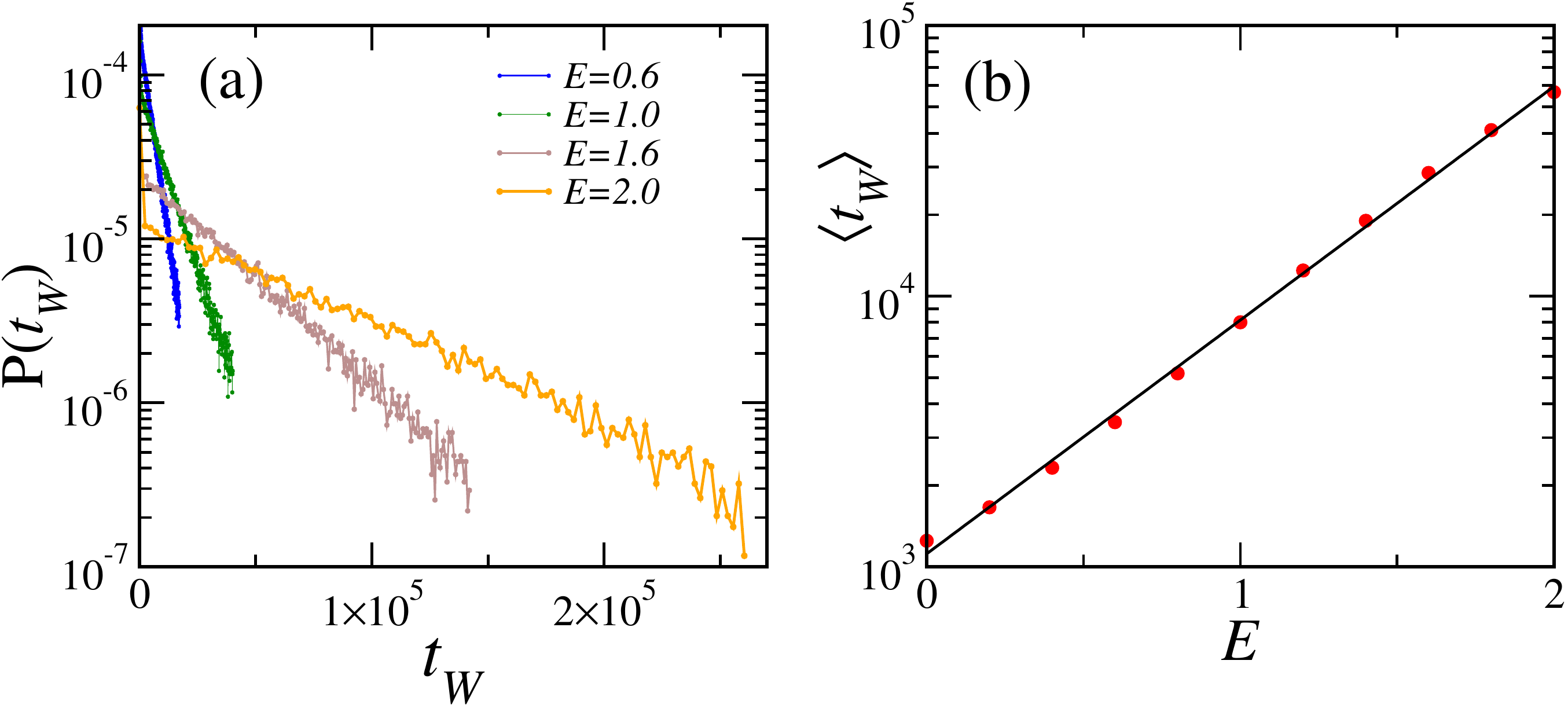}
 \caption{(Color online) (a) The probability distribution of the waiting time $t_{W}$ plotted in semi-log scale for different values of external force $E=0.6,1.0,1.6,2.0$ (from dark to light curves). (b) Average waiting time $\la t_{W} \ra$ of a particle versus external force $E$  
in a narrow tube with cages. The solid line corresponds to the best fit $e^{bE}$ with $b=1.988$. Here $X_L=Y_L=5, S_L= 2.5,\beta =1$ and $\gamma=1$.}
\label{fig:tube_pst}
\end{figure}

\noindent {\it Mapping to biased 1-d random walk.}\\
If we consider the cells of the narrow tube in Fig.~\ref{fig:tube}   as sites of a $1$-$d$ lattice, the motion of  the particle can be described as an effective biased random walk on this lattice. To check whether dynamical activity is still well-represented by the escape rates (such as for Markov processes), we measure the waiting time distribution of the particle in the cages. Let $t_{W}$ denote the waiting time of the particle in  the lower half of the cell. Fig.~\ref{fig:tube_pst}(a) shows  $P(t_{W})$ in the semi-log scale for different values of the driving force; it suggests an exponential probability density
\bea
P(t_{W} =\tau) =\lambda\,  e^{-\lambda \tau}
\eea
confirming the effective  Markov process picture. Here $\lambda= \frac{1}{\langle t_{W} \rangle}$ measures the escape rate from the cage.  Dependence of $\la t_{W}\ra$ on the external force $E$ for $\beta=1$ is shown  in semi-logarithmic scale in Fig.~\ref{fig:tube_pst}(b); the best fit straight line  is also added in the figure.  From the linear nature of this plot we infer,
\bea
\la t_{W}\ra  \sim e^{b(\beta)E} 
\eea
An  empirical study of $b(\beta)$ for different temperatures (not shown here) indicates that $b(\beta) \propto \beta.$ 
 The exponentially increasing average waiting time indicates the particle spends more and more time inside the cages as the external force is increased. The original $2$-dimensional nonequilibrium process  can then be thought of as an equivalent biased random walk on the $1$-dimensional lattice  with an escape rate 
\bea
g_\beta(E) = {1 \over \la t_W \ra} \sim e^{-b(\beta) E} \quad \text{with} \quad b(\beta) \propto \beta \n
\eea
which is indeed a decreasing function of the field strength.
This picture agrees with  the suggestions of Section \ref{sec:brw} --- a decreasing escape rate is a key ingredient of systems with negative differential response as that is mathematically picked up by the nonequilibrium response formula \eqref{eq:dQSD_cov} in the frenetic contribution.   \\
A fully discrete and Markovian version is discussed in Appendix \ref{app:dhooks}.

\begin{figure}[t]
 \hspace*{-1cm}\includegraphics[width=7 cm,bb=0 0 237 282]{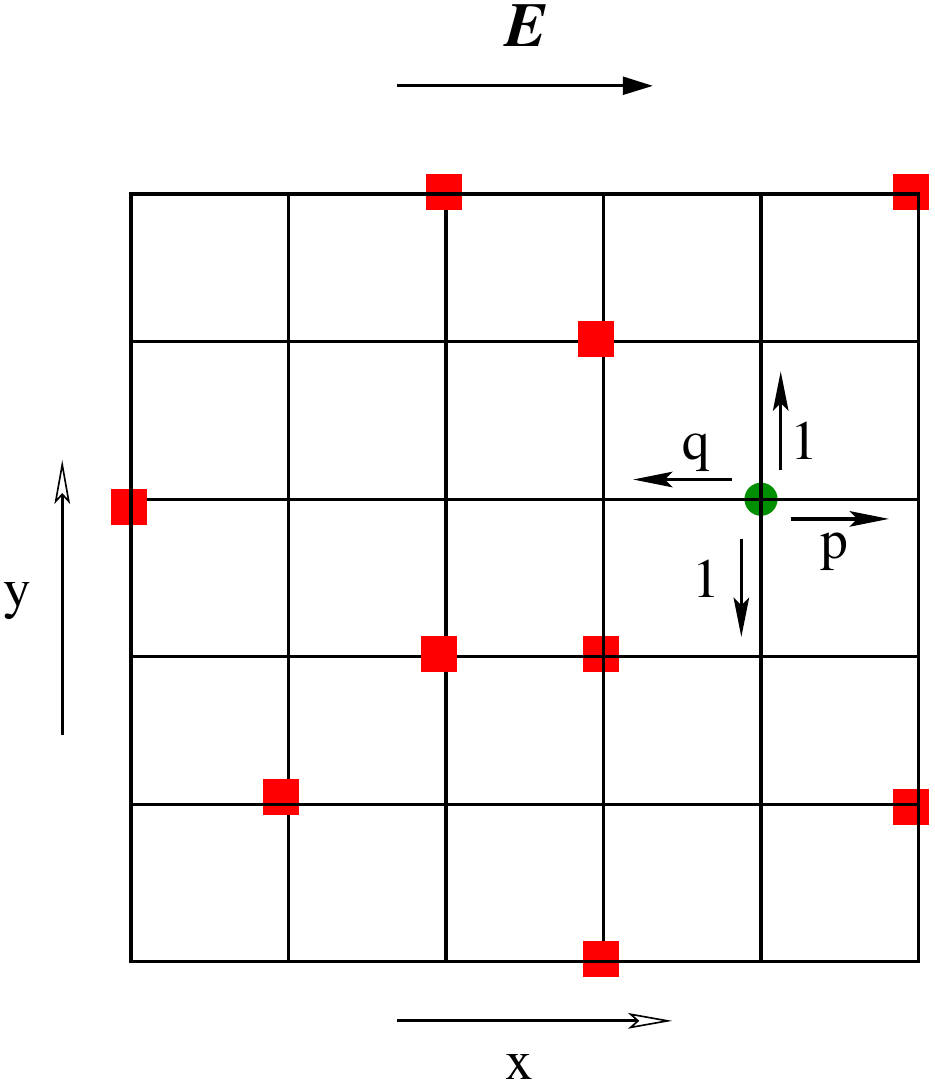}
 \caption{(Color online) Schematic diagram of the $2$-dimensional lattice Lorentz gas. The particle (shown as the green circle) performs a biased random walk; the red squares represent `obstacles' or inaccessible sites. }
 \label{fig:lorentz}
\end{figure}

\subsection{Driven lattice Lorentz model}\label{sec:Lorentz}

Our second example is the two-dimensional Lorentz gas \cite{Lorentz}, a well studied model of particle transport where a particle is allowed to freely diffuse in presence of random obstacles \cite{Beijeren,Dettmann}.\\
The field driven lattice Lorentz gas has been studied earlier in  the very wide context of diffusion in a random medium \cite{Barma} and it was shown that the drift velocity is a non-monotonic  function of the bias. In this section we investigate the origin of this non-monotonicity and following the main theme of the paper, we show that the presence of random obstacles  results in a decrease of the dynamical activity causing the negative mobility of the particle.\\
  
We consider a particle performing a continuous time two-dimensional random walk on a periodic square lattice of linear dimension $L$ where randomly a fraction $n$ of sites, called obstacles, have been made inaccessible.  Let us assume that the particle is driven in the $x$-direction by an external force field $E;$ local detailed balance suggests that $p/q= e^{\beta E}$ where $p(q)$ is the rate of moving forward (backward). That condition does not specify the individual rates fully but we choose $p=e^{\beta E/2}$ and $q = e^{-\beta E/2}.$ In the {\it absence} of obstacles such a choice corresponds to $g_\beta(E) \sim \cosh \beta E/2,$ where, as mentioned in Section \ref{sec:brw}, the Green-Kubo formula holds for all $E$ (no frenetic contribution at all.)  There is no bias in the $y$-direction and the rates of moving up and down are both assumed to be unity. However, the particle is blocked when the target site is inaccessible. Fig.~\ref{fig:lorentz} illustrates the set-up and dynamics.

\begin{figure}[t]
 \centering
 \includegraphics[width=16 cm]{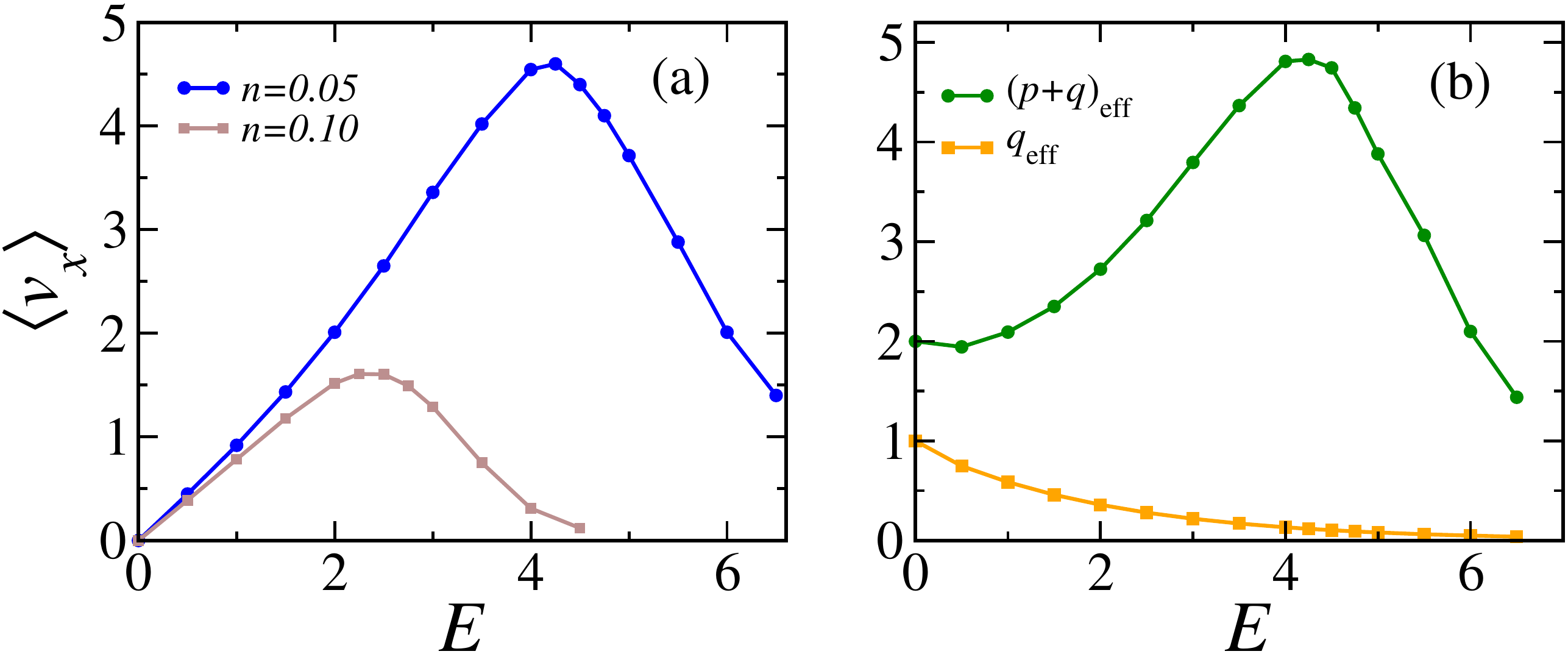}
 \caption{(Color online) Driven lattice Lorentz gas: (a) $\la v_x\ra$ {\it versus} $E$ plot for two different obstacle densities $n=0.05$ (dark blue circles)and $n=0.1$ (light brown squares). (b) Dependence of $(p+q)_{\mbox{eff}}$ (dark green circles) and $q_{\mbox{eff}}$ (light orange squares) on the external field $E$ for obstacle density $n=0.05.$ In both cases $\beta =1.$}\label{fig:Lorentz_v}
\end{figure}

We use numerical simulation to study the dependence of the average velocity $\la v_x \ra$ of the particle in the $x$-direction    on the field strength $E.$ Fig.~\ref{fig:Lorentz_v}(a) shows this plot for two different obstacle densities $n.$ The data are obtained by averaging over  at least $150$ obstacle configurations, with $100$ independent trajectories for each such configuration. 
The resulting curve shows a non-monotonic behavior, it decreases for large force $E$ after an initial increase consistent with the Green-Kubo formula. The decreasing velocity for large $E$ marks the negative differential mobility regime. As the obstacle density is increased the onset of the negative mobility shifts to smaller values of field $E.$  In contrast with the previous model of Section \ref{sec:hooks} the motion is left/right symmetric for $E=0$.   Moreover, there are no {\it a priori} constructed traps.  The trapping is more random and coming from obstacle configurations that make effective cages.

Here again we can follow the path-space approach of Appendix \ref{app:respa}  to understand the role of the frenetic contribution to the response of this system. For each trajectory $\omega$ over $[0,t]$  let
 $t_{RO}(t_{LO})$ denote the time during which there is an `obstacle' at the right(left) neighboring lattice site of the  particle.   Then, \eqref{eq:ShDh} gives
\bea
S(\omega)& =& (N_\rightarrow - N_\leftarrow) \log \frac pq = JE\cr
D(\omega)& =& p\,(t-t_{RO}) + q\,(t-t_{LO})
\eea
As before the perturbation considered is a small increase in the external field $E \to E+\id E.$
The linear response relation for any observable $O$ is then written following Eq. \eqref{eq:dQSD},
\bea
\frac \id{\id E} \la O\ra^E = \frac \beta2 [ \la J O\ra -(p-q)t \la O\ra + p \la t_{RO} O\ra -q \la t_{LO} O\ra]\cr \label{eq:dJdF}
\eea
This formula holds true for any initial configuration of the system and therefore can be applied in both 
transient and stationary regimes. We are particularly interested in the linear response of the velocity $ v_x  = J /t$ in the large $t$ limit,


\bea
\frac \id{\id E} \la v_x\ra^E = \frac {\beta t}{2} \la v_x;v_x \ra + \frac \beta2 [p \la v_x;t_{RO} \ra - q \la v_x;t_{LO} \ra] \n
\eea
\begin{figure}[t]
 \centering
 \includegraphics[width=16 cm]{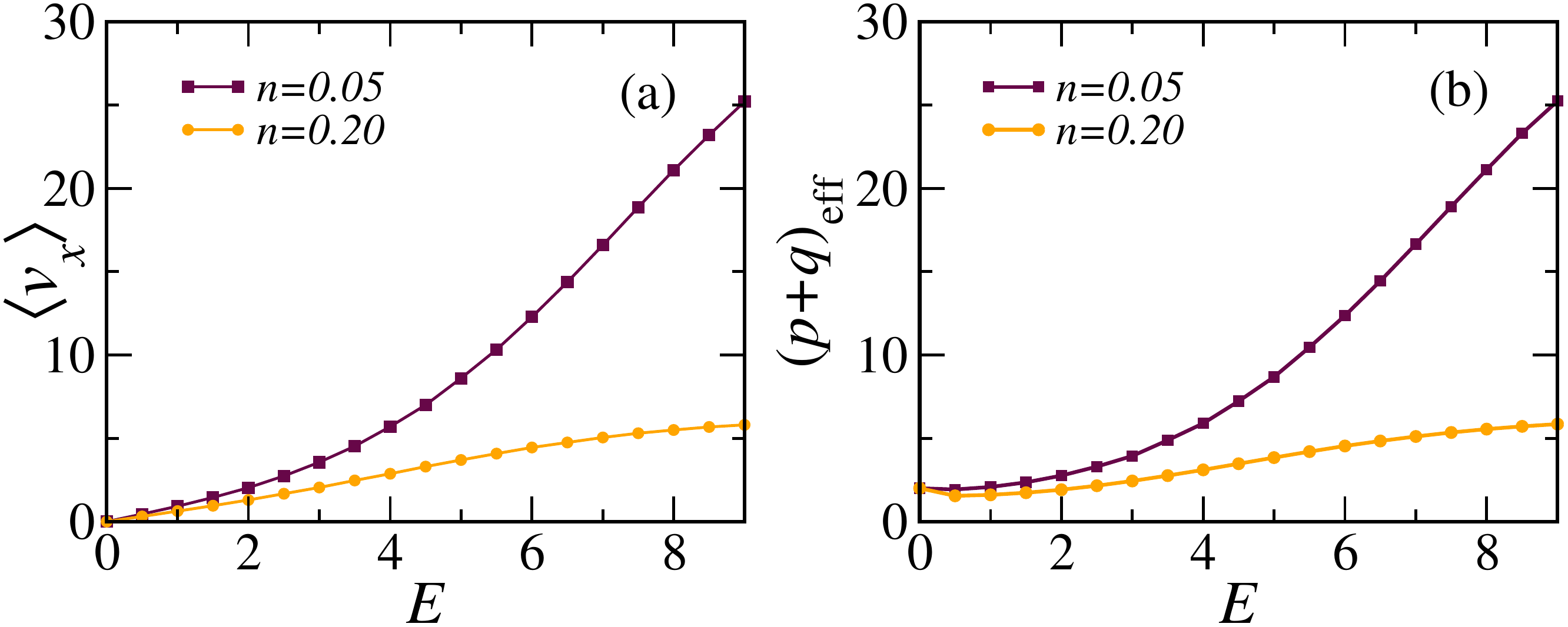}
 \caption{(Color online) Lorentz gas with obstacles in $x$-directions only: (a) The average velocity $\la v_x \ra$ increases monotonically with the driving field $E.$ The two curves correspond to two different obstacle densities $n=0.05$ (dark maroon squares) and $n=0.20$ (light orange circles). (b) The corresponding escape rates $(p+q)_{\text{eff}}$ are also increasing functions of the field $E$ indicating there is no `trapping' in this case.}\label{fig:x_obstacle}
\end{figure}
The first term, in the limit of $t \to \infty,$  is proportional to  the  diffusion constant $D_\text{dif}$  for the particle and is always positive. The observed negative mobility  can only be caused by the second term,  for example when $v_x$ and $t_{LO}$ are highly positively correlated.\\
If we take a constant $O=1$ in \eqref{eq:dJdF}, then the left-hand side vanishes which gives us a relation between the stationary state current $J$ in the $x$-direction and the rates,
\bea
\la J \ra &=& (p-q)t - p \la t_{RO}\ra + q \la t_{LO}\ra \cr
&=& (p_{\mbox{eff}} - q_{\mbox{eff}})t \label{eq:J_t_Lorentz}
\eea
where we have defined
\bea
p_{\mbox{eff}} = p\left(1- \frac {\la t_{RO} \ra} t\right) \; \text{and} \; q_{\mbox{eff}} = q\left(1- \frac {\la t_{LO} \ra} t \right) \n
\eea
This relation allows us to map the dynamics of the lattice Lorentz gas to that of an effective biased $1$-d random walker with rates $p_{\mbox{eff}}$ and $q_{\mbox{eff}}.$  In other words we are back to the biased random walker of Section \ref{sec:brw}.  The sum $(p+q)_{\mbox{eff}} \equiv p_{\mbox{eff}} + q_{\mbox{eff}} = g_\beta(E)$ plays the role of effective escape rate from a site. Unsurprisingly, $(p+q)_{\mbox{eff}}$ is non-monotonic in the field strength $E,$ as shown in Fig. \ref{fig:Lorentz_v}(b) and the conclusions of Section \ref{sec:brw} apply.\\

At the end of the previous section we mentioned that the presence of obstacles in both the $x$ and $y$ directions are crucial for the trapping of the particle. In the case of the Lorentz gas we can see this immediately by studying a variation where the obstacles do not block the motion in the $y$ directions. Numerical simulations show that the system does not show any negative mobility in this case; the stationary velocity is a monotonically increasing function of the external field.  Figure \ref{fig:x_obstacle}(a) shows current versus field for densities $n=0.05, 0.2.$ In agreement with our claim, the $(p+q)_{\mbox{eff}}$ is a monotonically increasing function in this case (Fig. \ref{fig:x_obstacle}(b)).  
So it is not just the fact that there are obstacles; it is the caging effect which is important.

\section{Thermal transport}\label{sec:thertr}

As a second major case we look here at thermal conductivity.  We ask how the transport of thermal energy is affected 
when  some ambient temperature is changed. In this section we give a scenario for  negative differential thermal conductivity, which again will be traced back to the frenetic contribution.

Thermal conductivity (or resistivity) measures the change in the current when the magnitude of the thermal gradient is changed. Close-to-equilibrium thermal conductivity is a positive quantity. Here we are interested in systems which show the counter-intuitive property of negative differential thermal resistance (NDTR), a decrease in thermal current when the temperature difference between the two ends of the system is increased.  In recent years there have been several studies \cite{Li,Yang,Hu,Shao,LiRMP} where NDTR has been observed by various nonlinear mechanisms.  We believe they are all related more specifically to the negative frenetic contribution, which we make explicit in a simpler model.\\

 \begin{figure}[t]
 \centering
 \includegraphics[width=11 cm]{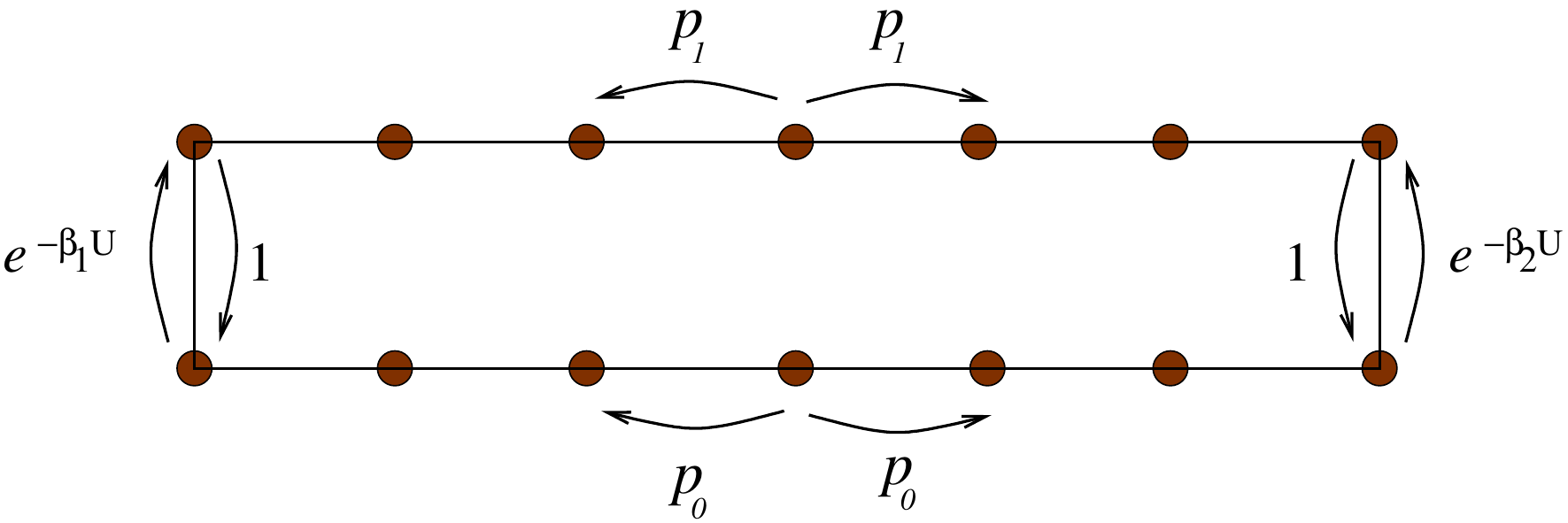}
 \caption{(Color online) Schematic representation of the discrete model for heat conduction.  The horizontal direction is spatial, the vertical direction is energetic.  Heat exchange is only possible at the edges.}
 \label{fig:4stmodel}
\end{figure}

Let us consider  $L$ consecutive sites labeled by $i=1,\dots,L.$ Associated with each site $i$ are two states carrying different energies. As shown in Fig.~\ref{fig:4stmodel}, one can think of a two-lane model; the lower lane and upper lane carry energies $U_0$ and $U_1$ respectively. Energy quanta are hopping symmetrically along these lanes without inter-lane transitions. The system is allowed to exchange energy with the environment only at the left and right edges where it is attached to two heat baths of temperatures $T_1$ and $T_2$ respectively. We denote the  state of the system by $x^{u,d}_i$ where $u,d$  refer to the upper/lower  energy lanes. The dynamics is then completely specified by the following rates,
\bea
k(x^d_1,x^u_1)&=& e^{-\beta_1 U}, \quad k(x^u_1,x^d_1) = 1 \cr
k(x^d_L,x^u_L)&=& e^{-\beta_2 U}, \quad k(x^u_L,x^d_L) = 1 \cr
k(x^d_i,x^d_{i\pm1})&=& p_0 , \quad \;\;\; k(x^u_i,x^u_{i\pm1}) = p_1
\eea
Here $\beta_{1,2}$ are the respective inverse temperatures of the left and right baths and $U=U_1-U_0$ is the energy difference between the two lanes. Without any loss of generality we assume energies $U_0=0$ and $U_1=U.$

Let $N^{u,d}_{\rightleftarrows}$ denote the total number of jumps to the right and left in the upper and lower lane. Similarly $N^{l,r}_{\uparrow\downarrow}$ denote the number of jumps to the upper and lower levels at the left and right bonds. The heat or energy transported  through the system over a time $[0,t]$ is given by
\bea
J &=& U_1 (N^u_\rightarrow -N^u_\leftarrow ) +U_0 (N^d_\rightarrow  -N^d_\leftarrow ) \cr
&=& U (N^u_\rightarrow  -N^u_\leftarrow )
\eea

We assume $T_1>T_2$ so that the system is expected to have a constant heat or energy  current $\la J\ra$ flowing  
from left bath to right one in the stationary state. Near equilibrium i.e., when the temperature difference $\Delta T = T_1-T_2$ between the two baths is small this current is proportional to $\Delta T $ (Fourier's law) no matter how we choose $p_0,p_1$.  For large gradient that need not be true.
Suppose indeed that we introduce a temperature dependence in the symmetric jump rate $p_0=T_1 T_2$ which decreases as the temperature of the cold bath $T_2$ is decreased;  $p_1$ is taken independent of temperatures.  That provides a trapping mechanism for the system in the lower lane configurations $x_1^d$ and $x_L^d$.  Other set-ups are possible but the main idea is to let kinetic factors of transport be negatively influenced by lowering one of the edge-temperatures.\\  

\begin{figure}[t]
 \centering
 \includegraphics[width=16 cm]{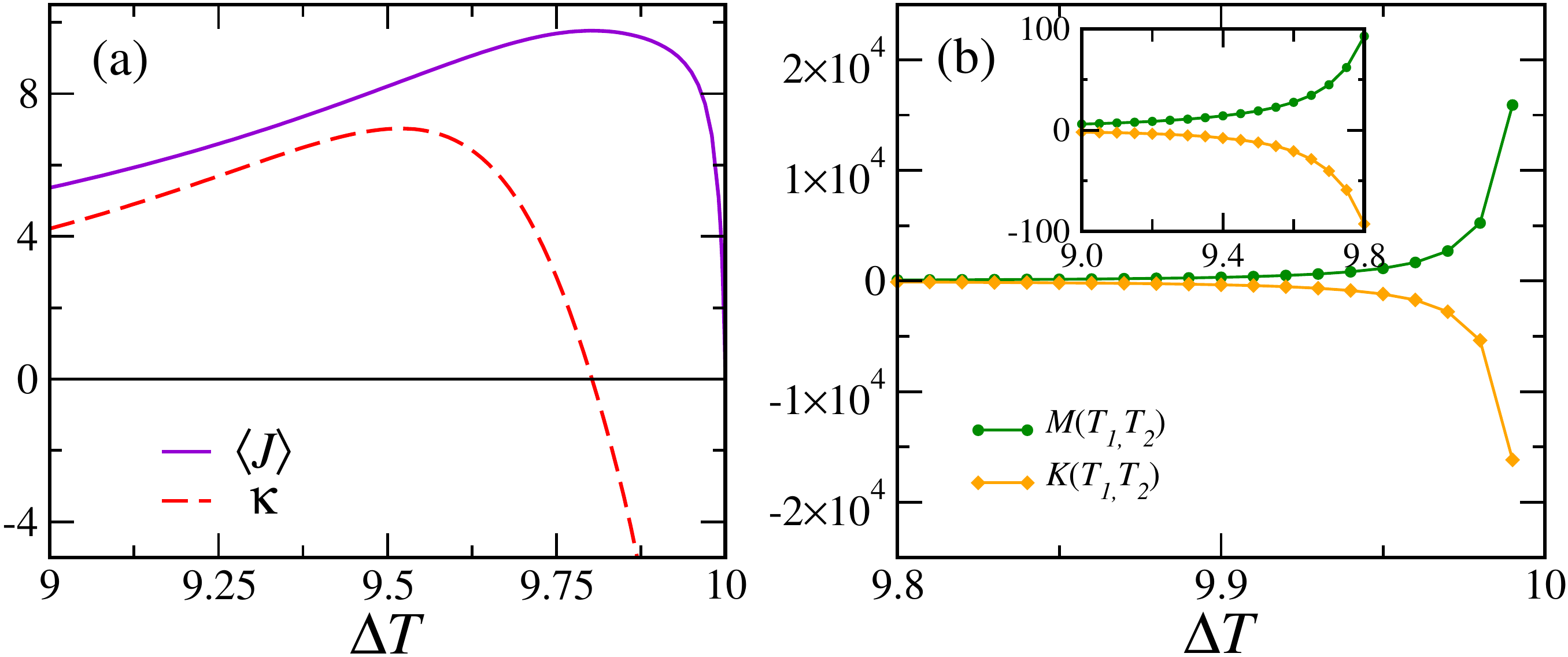}
 \caption{(Color online) Thermal conductivity: (a) Both current $\la J\ra$ (solid line) and the conductivity $\kappa$ (dashed curve) as a function of the temperature difference $\Delta T.$  (b) The entropic (dark green circles) and frenetic (light orange diamonds) components of the response function $\kappa$. These curves added together result in the dashed curve of (a).  The inset shows the same for a different range of $\Delta T.$ Here the hot bath is fixed at temperature $T_1=10.0,$ $p_0=T_1T_2,$ with $T_2=T_1-\Delta T.$ The other parameters are $p_1=1.0,U=1.0$.   The time interval has $t=100$ and the data are averaged over $10^7$ independent ensembles.}
 \label{fig:J_T}
\end{figure}

The simplest case is when $L=2$, in which case we have only $4$ sites. The results of the simulation are shown for that case (where it is also possible to exactly calculate the average current) but the result remains entirely similar when longer systems are considered. The dependence of the thermal current $\la J\ra$ on the temperature difference $\Delta T$ is shown in Fig.  \ref{fig:J_T}(a) for $T_1=10.0$ (solid line), though initially increasing, the current drops down as $\Delta T$ approaches $ T_1$, i.e., as $T_2 \to 0,$ marking a negative differential thermal response. 

The rate of change of thermal current with the temperature difference $\Delta T$ between the two baths is
\bea
\kappa \equiv {\id \la J \ra \over \id \Delta T} = \frac 12 \left\la J; {\id \over \id \Delta T} S(\omega) \right\ra^{T_1,T_2}-\left\la J; {\id \over \id \Delta T} D(\omega) \right\ra^{T_1,T_2} \n
\eea
The last equation follows from Eq. \eqref{eq:dQSD_cov} where $\Delta T$ acts as the driving field, with $T_2=T_1-\Delta T.$ The entropy $S(\omega)$ and dynamical activity $D(\omega)$ associated with a path $\omega$ are obtained following \eqref{eq:ShDh},
\bea
S(\omega) &=& (N^l_\downarrow - N^l_\uparrow) \beta_1 U + (N^r_\downarrow - N^r_\uparrow) \beta_2 U \cr
D(\omega) &=& -(N^u_\rightarrow + N^u_\leftarrow) \log p_1  - (N^d_\rightarrow + N^d_\leftarrow) \log p_0 \cr 
&+&  \frac 12 (N^l_\downarrow + N^l_\uparrow) \beta_1 U  + \frac 12 (N^r_\downarrow + N^r_\uparrow) \beta_2 U +  \sum_{i=1 \atop \alpha =u,d}^L \xi^\alpha_i  t^\alpha_i \n
\eea
where the $t^\alpha_i$ are the residence times of states $x^\alpha_i$ and the $\xi$'s are the corresponding escape rates.

The entropic component of the thermal conductivity can be calculated from the above equations,
\bea
M(T_1,T_2) &\equiv& \frac 12 \left\la J; {d \over d \Delta T} S(\omega) \right\ra ^ {T_1,T_2} \cr
&=& \frac{U}{2T_2^2}\la J;(N^r_\downarrow- N^r_\uparrow  ) \ra = \frac{1}{2T_2^2(L-1)}\la J;J \ra ~~~~~~
\eea
In the last equality  we have assumed the large time limit.
As always, this term is positive definite and gives the Green--Kubo formula in equilibrium. The other component, arising from the  correlation with dynamical activity, comprises of several contributions. For the simplest case $L=2$ it has the form
\bea
K(T_1,T_2) &\equiv& -\left\la J; {d \over d \Delta T} D(\omega) \right\ra \cr
&=&  -\frac 1{T_2} \la J;(N^d_\rightarrow + N^d_\leftarrow ) \ra - \frac {U}{2T_2^2} \la J;(N^r_\uparrow + N^r_\downarrow)\ra \cr &&  + T_1\, \la J;t^d_1\ra + \left(T_1 + \frac U{T_2^2} e^{-\frac U{T_2}}\right) \la J;t^d_2 \ra 
\eea
The first two terms quantify the correlation of the current with the total number of jumps in the lower and right bonds, whereas the two last terms contain the correlation with the time spent in the configurations $x_1^d$ and $x_2^d$ in the lower lane. Fig. \ref{fig:J_T}(b) shows separate plots of the quantities $M(T_1,T_2)$ and $K(T_1,T_2);$  the frenetic component  shows large negative  contribution. In fact, though the two curves look like mirror image of each other they do differ on a much smaller scale.

\section{Additional remarks}\label{sec:rema}

\begin{enumerate}
\item The origin of negative differential response need not always be frenetic.  A  more complete and correct (but also more complicated) title of the present paper would be ``The time-symmetric origin in nonequilibrium ensembles for the negative differential response in time-antisymmetric variables.''
 Not considered in the present paper but still interesting response indeed deals with time-symmetric observables, e.g.  for the dynamical activity itself or  for time-symmetric currents as occur with momentum transfer.  The situation then gets reversed with respect to the present study.  At equilibrium the Green-Kubo relation would be reconstructed from the correlation of the observable with the dynamical activity, and nonequilibrium corrections would be entropic.  At equilibrium there is no real distinction.\\

\item Note that the negative differential response sets in at intermediate values of the (driving) field $h$, not necessarily very large.  In fact, it is also possible to observe the same effect of negative differential response at intermediate driving while the current starts to increase again for large values of the driving. In particular the current does not need to vanish for large external field. As an example one can consider the model discussed in Section \ref{sec:hooks}, but with `soft hooks' which can be crossed with a small probability.  If we include a small rate of crossing the barriers, then after an initial increase, the current drops marking the negative conductivity regime,   
but at large biasing field the current rises again.\\

\item There are by now various mathematically equivalent formulations of linear response in nonequilibrium; see the review in \cite{update}.  They do not however appear  equally useful in all circumstances.  We feel that for a unifying framework of negative response, the one starting from the path-integration reviewed in Appendix \ref{app:respa} is most promising.  It remains however interesting to relate the present approach with for example ideas around negative effective temperature.  Let us take for simplicity the biased random walker of Section \ref{sec:brw}.  Equation \eqref{eq:dJdE_rw} can be rewritten as proportional to the current-current correlation
\begin{equation}
{\id \over \id E}\la J\ra =\frac{\beta_{\mbox{eff}}}{2} \la J;J \ra
\end{equation}
simulating the Green-Kubo expression but with effective temperature given by
\[
\beta_{\mbox{eff}} = \beta + 2\left( \frac {g^\prime_\beta(E)}{g_\beta(E)}- \frac \beta 2 {1- e^{-\beta E} \over 1+ e^{-\beta E}}  \right) {1- e^{-\beta E} \over1+ e^{-\beta E}}
\]
Clearly, a negative differential conductivity is accompanied by a negative $\beta_{\mbox{eff}}.$ The actual dependence of the effective temperature on the external field $E$ depends on the escape rate $g_\beta(E)$.  $\beta_{\mbox{eff}}$ is  shown as a function of $E$ for two different temperatures and $g_\beta(E) = (1+ (\beta E)^2)^{-1}$ in Fig.~\ref{fig:beff}.  There, $\beta_{\mbox{eff}} \to 0$ for large $E$.  When for large $E$, $g_\beta(E) \sim e^{-\alpha \beta E}$, then $\lim_{E \to \infty} \beta_{\mbox{eff}} \to -2\alpha \beta.$


\begin{figure}[t]
 \centering
 \includegraphics[width=8 cm]{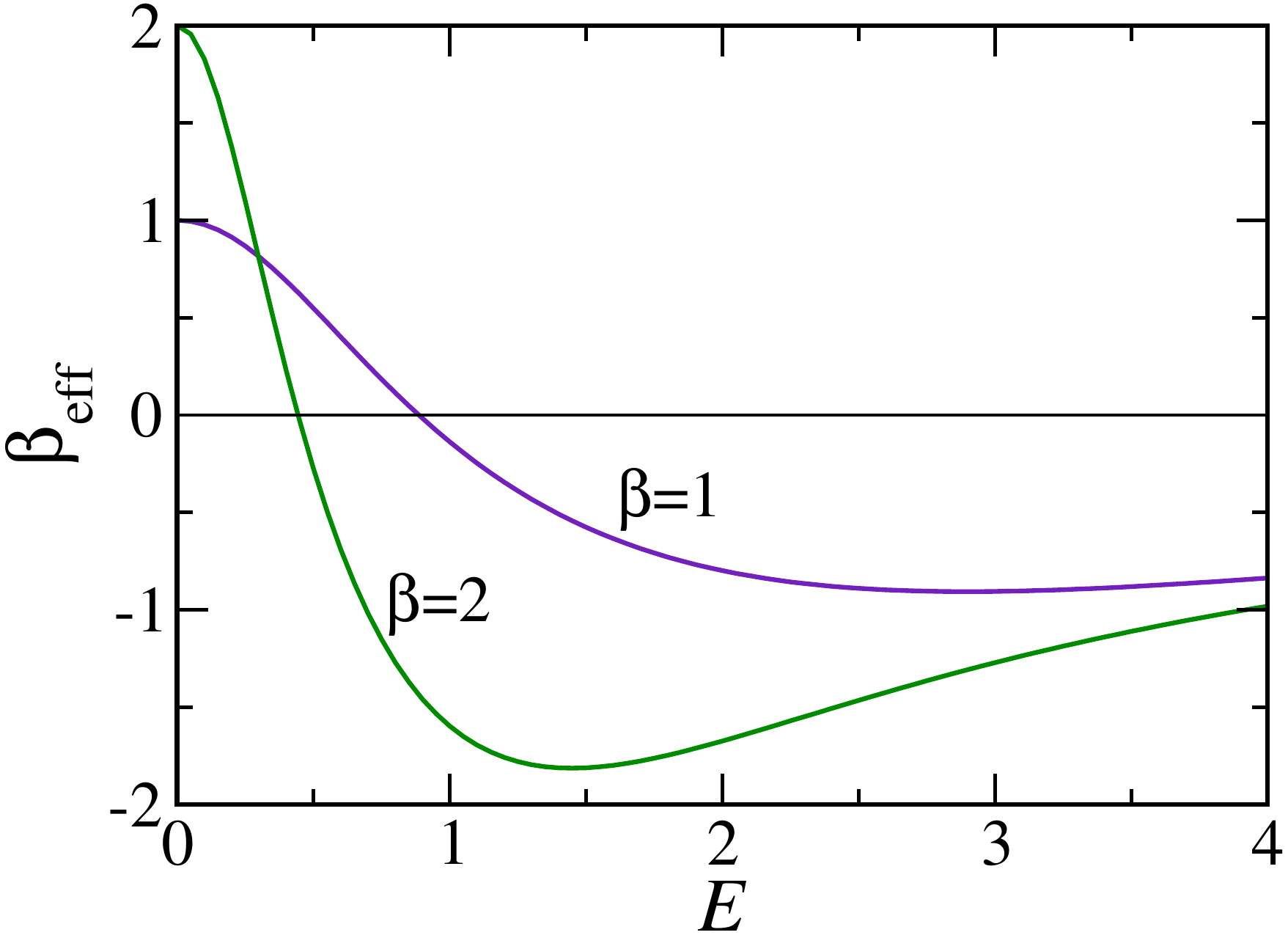} 
 \caption{(Color online) Variation of the effective temperature $\beta_{\text{eff}}$ with the external field $E$ for $\beta=1,2$ and for $g_\beta(E) = (1+ (\beta E)^2)^{-1}.$ }
 \label{fig:beff}
\end{figure}

\item
There are other aspects of negative response which fall outside the discussion of the present paper.  That is for example the case for the occurrence of negative heat capacities in nonequilibrium multilevel systems, \cite{negcap,jirka}.  It has not yet been sufficiently understood how to identify there the origin of negative (thermal) response in terms of the frenetic contribution.\\

\item As frenetic effects make it possible to have negative differential response, they are also the cause of having zero differential response, for example at the (temperature dependent) field value $E^*$ in the model \ref{sec:hooks}.  Considering the model exactly at that value, there is no linear response and the change in current  $\langle J \rangle^{E^*+\id E} - \langle J \rangle^{E^*}\propto (\id E)^2$ starts off nonlinearly in $\id E$.

\end{enumerate}

\section{Summary and general discussion}

We have discussed a general formalism to understand negative differential responses in far from equilibrium systems. The prototypical example of a biased random walker where the escape rates are field dependent already makes the point quite clearly.  When kinetic factors such as trapping mechanisms, collision frequencies, reactivities etc. are dependent on the nonequilibrium driving, they get a strong influence on the response via the frenetic contribution.  We have seen that both in particle and thermal transport.\\
\begin{figure}[t]
 \centering
 \includegraphics[width=8 cm]{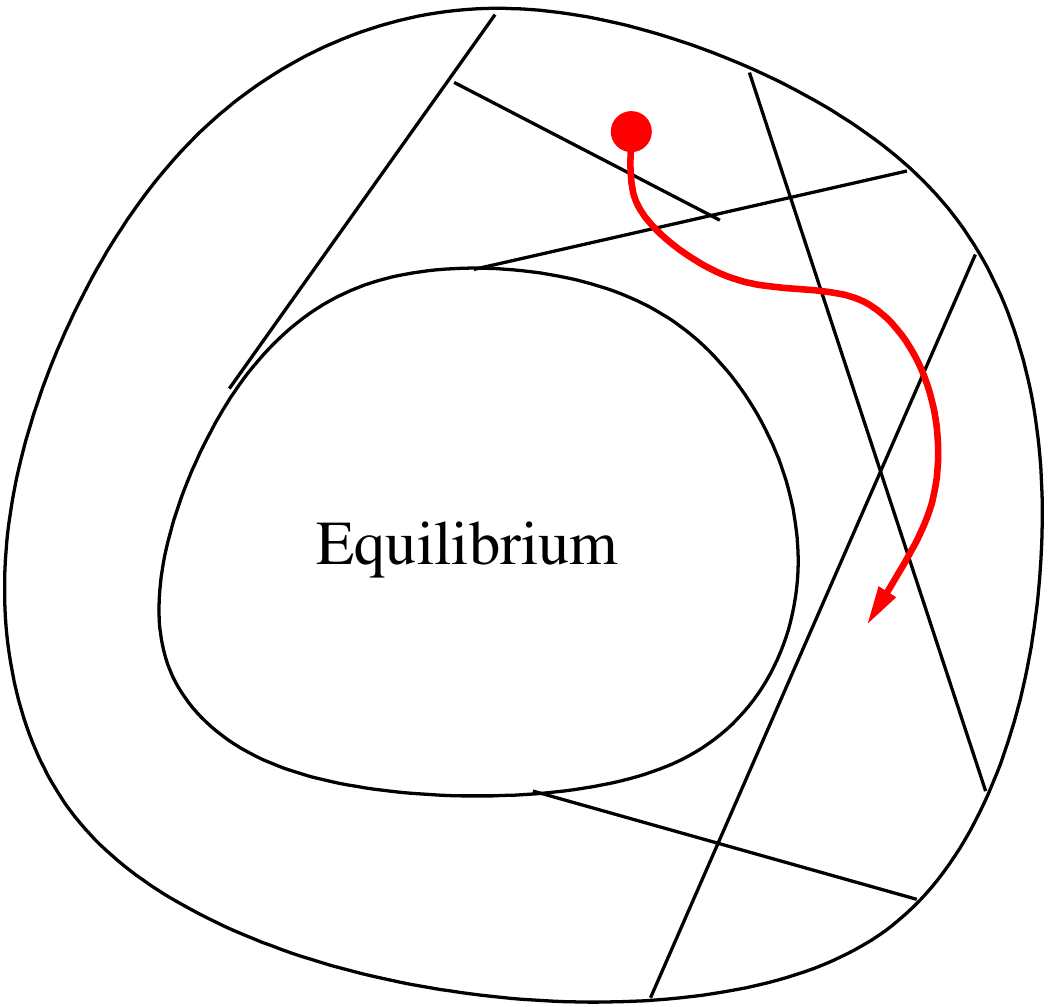}
 \caption{(Color online) Cartoon of the very high-dimensional classical phase space for a macroscopic mechanical system. Each region corresponds to a reduced description or physical coarse-graining, say collecting all microscopic states that correspond to a particular position of a tagged particle and a certain energy and particle number in each of the reservoirs.  For the time-scale of nonequilibrium phenomena the trajectory of the microscopic state visits much smaller regions of phase space as compared to equilibrium.}
 \label{fig:boltz}
\end{figure}

  To lift the discussion to some more general phase space considerations, we would like to remind the reader of the phase space picture in Fig.~\ref{fig:boltz}.  We see the usual state space of a mechanical system where each point collects the information of the positions and momenta of all the many particles.  Say for the motion of colloids in the narrow tube as discussed under Section \ref{sec:hooks}, the mechanical system is the closed and isolated system containing both reservoirs (heat and particle baths organizing the isothermal driving) and colloids.  We look over time-scales where the nonequilibrium condition exists (before any global relaxation to equilibrium is apparent).  The phase space is divided in regions that each collect all states of the mechanical system that correspond to certain positions of the colloids and to certain values of the energy and particle number in the reservoirs.  The biggest region (in terms of volume or entropy) is the equilibrium situation.  Under nonequilibrium the 
mechanical 
trajectory is visiting regions in phase space that are tiny (in volume-sense) compared to equilibrium.   The dynamics now runs effectively
   between relatively small phase space volumes.  At that moment, not only the volume (read: entropy) but also surface considerations start to matter.  The surface-area measures the interface between different phase space regions in terms of exit and entrance rates, for short the dynamical activity as we have discussed in the present paper.  Negative differential response then corresponds to kinetic constraints or caging effects restricting mechanical motion between different phase space regions.\\

\noindent {\bf Acknowledgments:} We thank Marco Baiesi for initial discussions that also have lead to the formulation of the example \ref{sec:brw} in Section II.  We are also grateful to Abhishek Dhar for suggesting a number of relevant references.

\appendix

\section{Response from path-integration}\label{app:respa}

Dynamical ensembles in nonequilibrium statistical mechanics are represented by a probability measure ${\c P}(\omega)$ on path space.  This measure depends on the parameters of driving and reservoirs and would generally change when a perturbation is added to the system. Let us think of a generic perturbation $h \to h+\id h$ which changes the probability measure ${\c P}^h(\omega)$ to ${\c P}^{h+\id h}(\omega).$ We compare
the path weights with a reference process and associate an action $A(\omega)$ to each trajectory $\omega$ via ${\c P}(\omega)=e^{-A(\omega)}{\c P}_0(\omega)$ where  ${\c P}_0(\omega)$ is the weight of the same path for the reference process. 

The change in expectations for an observable $O$ due to the perturbation is now conveniently expressed as,
\bea
&& \la O(\omega) \ra^{h+\id h} -\la O(\omega) \ra^{h}  \cr &&=  \int d\omega {\c P}_0(\omega ) 
\left(e^{-A_{h+\id h}(\omega )}-e^{-A_{h}(\omega )}\right) O(\omega ) \n
\eea
For small perturbations $\id h$ this leads to a general differential response formula \cite{update},
\bea
\frac \id{\id h} \la O(\omega )\ra^h = - \left \la O(\omega )\frac {\id}{\id h} A_h(\omega )  \right\ra^h \label{eq:dQA}
\eea
where the right-hand side is an average over the unperturbed process. 

It is useful to decompose the action into two components by writing $A_h(\omega ) = D_h(\omega ) - \frac 12 S_h(\omega ),$ where $S_h(\omega )$ is the time anti-symmetric entropy associated with the trajectory $\omega $ and the time-symmetric part is the dynamical activity $D(\omega )$\, \cite{fdr}. The response relation \eqref{eq:dQA} now takes the form \eqref{eq:dQSD}.\\

To apply this formula to specific systems one needs to determine $S_h(\omega )$ and $D_h(\omega ).$ Let us derive the formul{\ae} \eqref{eq:ShDh}
mentioned for Markov jump processes; see \cite{fdr,markov} for more details.\\
Let the transition rates between states $x\rightarrow y$ be $k(x,y)$.  Escape rates are $\xi(x) = \sum_y k(x,y)$.  Paths $\omega$ are piece-wise constant with jumps at times $s_i$ and have weight
\bea
{\c P}^h(\omega ) = \mu_0(x_0)\prod_{s_i} k(x_{s_i},x_{s_{i+1}}) e^{-\int_0^t \xi(x_s) \id s}\n
\eea
for initial distribution $\mu_0(x_0).$ To write the action $A_h(\omega )$ we need to choose a reference process. It is easy to show that the final response formula does not depend on this choice. So, for our purpose we take the simplest reference process defined by $k_0(x,y)=1$ iff $k(x,y) \ne 0.$  Then, 
\bea
A(\omega ) = -\sum_{s_i} \log k(x_{s_i},x_{s_{i+1}}) + \int_0^t ds[\xi(x_s) -\xi_0(x_s)]\cr \label{eq:A}
\eea
 The entropy and dynamical activity associated with trajectories can be identified as the time anti-symmetric and symmetric components of $A(\omega ).$ Denoting the time--reversed trajectory as $\theta \omega $,
\bea
S_h(\omega) &=& A(\theta \omega ) - A(\omega )  \cr
D_h(\omega ) &=& \frac 12 [A(\theta \omega ) + A(\omega )] \label{eq:ShDhbis}
\eea
from which \eqref{eq:ShDh} follows.  Note that $\int \id s\, \xi_0(x_s)$ in \eqref{eq:A} can be ignored for differential response as it does not depend on $h$.

\section{Discrete hooks}\label{app:dhooks}
Another discretization of the model in Section \ref{sec:hooks} is to define a Markovian random walker in one dimension following Fig.~\ref{fig:cm}.  A single particle walks in a long channel consisting of identical cells; each cell is again divided into $4$ parts labelled $i=1\dots 4$. There is a field $E$ in the horizontal direction which creates a bias in the rates of moving forward and backward, but motion in the vertical direction is unbiased. The corresponding rates can be expressed as
\begin{eqnarray*}
k_\rightarrow &=& e^{\beta E/2}\qquad   k_\leftarrow  = e^{-\beta E/2}\\
k_\uparrow &=&  1\;\;\;\qquad \;\;\, k_\downarrow = 1.
\end{eqnarray*} 
 
\begin{figure}[t]
\centering
\includegraphics[width=11 cm]{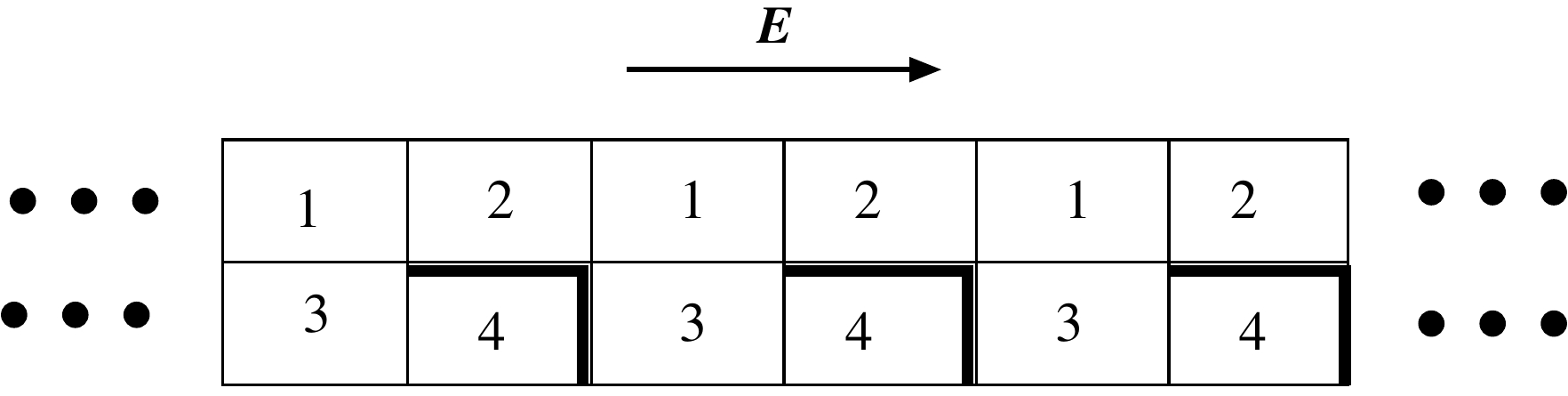}
\caption{Schematic diagram of the discrete model with cages.}\label{fig:cm}
\end{figure}

  A hard wall prohibits jumps between parts $2$ and $4$ of the same cell and from $4$ to $3$ of the next cell in forward direction. This model is basically the one studied in \cite{Zia} except for the fact that here the hard walls are placed at regular intervals.  We assume periodic boundary conditions in the horizontal direction.

\begin{figure}[t]
\centering
\includegraphics[width=16 cm]{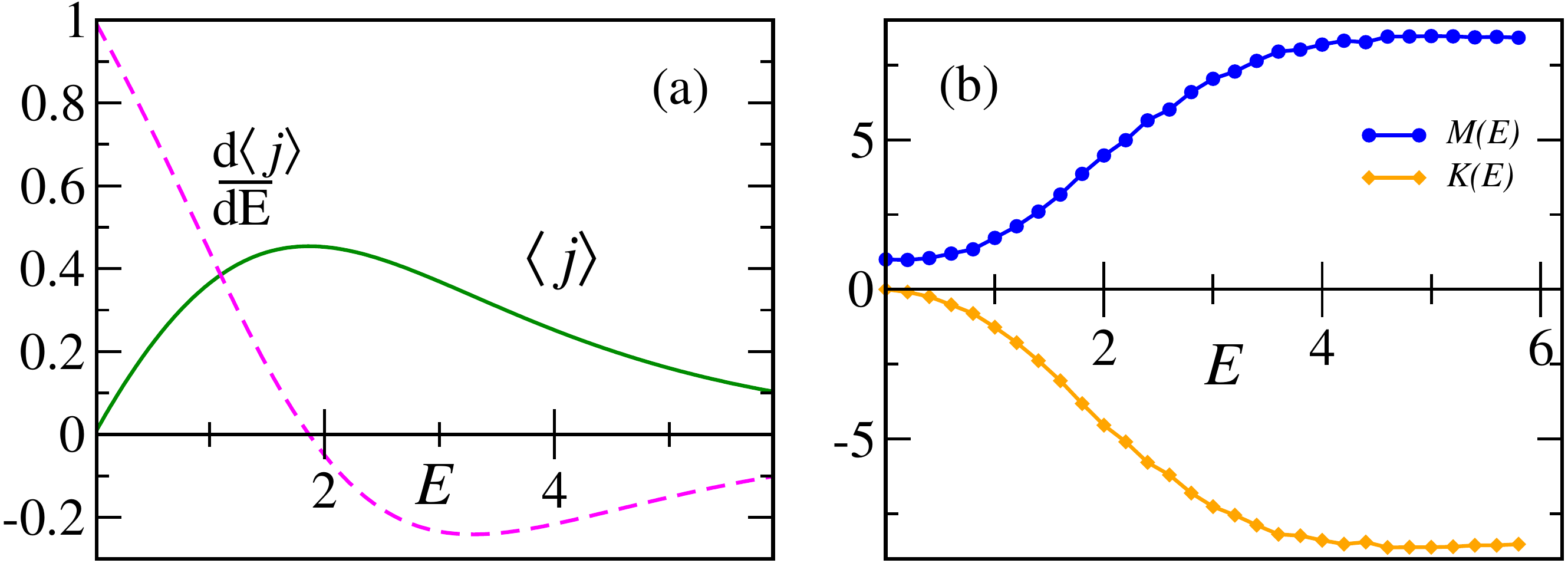}
\caption{(Color online) Discrete model with cages: (a) The current per unit time $\la j\ra$ (solid line) and conductivity $\frac {d \la j \ra}{dE}$ (dashed line) similar to Fig.~\ref{fig:tube_vel}(a).  (b) Entropic (dark blue circles) and frenetic (light orange diamonds) contributions to the response of $j$ as a function of $E$ following Eq. \eqref{eq:MK}. The time interval $t=10^4$ and the data are averaged over $10^7$ independent ensembles.}\label{fig:ST}
\end{figure}

It is straightforward to calculate the stationary current by solving the corresponding master equation.
Fig.~\ref{fig:ST}(a) shows the current (solid line) as a function of field strength $E;$ for convenience we have plotted  $\la j\ra= \la J\ra/t.$ After an initial increase the current decreases for large field and eventually vanishes: the upper sites, which contribute to the current, become exponentially less likely to be populated as $E$ is increased which overcompensates  the increasing bias in the forward rate.  Instead of giving the analytic solution we concentrate again on the response formula to find that the negative differential mobility can be attributed to the frenetic contribution.

The observed quantity is, once again, the average current in the forward direction over a time interval $[0,t]$, $\la J \ra =\la N_\rightarrow - N_\leftarrow \ra$ with $N_\rightarrow$ and $N_\leftarrow$ the number of jumps in the forward and backward directions respectively. Let $t_i$ be the time spent during a trajectory $\omega$ by the particle in the $i^{th}$ site; $\sum_i t_i = t.$  The escape rates are
\bea
\xi(1) &=& 1+ e^{\beta E/2} + e^{-\beta E/2} \quad \xi(3)= 1+ e^{\beta E/2}  \cr
\xi(2) &=& e^{\beta E/2} + e^{-\beta E/2} \quad \quad \quad \xi(4) = e^{-\beta E/2}
\eea

The entropy and dynamical activity associated with the path takes the simple forms,
\bea
S(\omega ) &=& (N_\rightarrow -N_\leftarrow)E \cr
D(\omega ) &=& \sum_{i=1}^4 \xi(i) t_i
\eea
where $t_i$ is the total time the particle spends in the $i^{th}$ part over the time-interval $[0,t].$

Finally, using \eqref{eq:dQSD} the response can be expressed as a sum of the correlations with excess entropy and excess activity.
\bea
\frac{d}{dE} \la J \ra^E &=& \frac 12 \left\la J ; \frac d{dE} S(\omega ) \right\ra^E -\left\la J ; \frac d{dE} D(\omega ) \right\ra^E \cr
&=& \frac 12 \la J;J \ra^E + e^{-\beta E/2} \langle(t -t_3);J\rangle ^{E} \cr 
&& \; - e^{\beta E/ 2} \langle (t - t_4);J\rangle ^{E} \n
\eea

We stick to the velocity $j=J/t$ and, after a small calculation, obtain
\bea
\frac{\id}{\id E}\la j \ra^E &=& M(E) + K(E), \;\;\text{with} \cr
M(E) &=& \frac 1{2t}\la J;J \ra^E \cr
K(E) &=& e^{-\beta E / 2} \langle(t -t_3);j\rangle ^{E}- e^{\beta E / 2} \langle (t - t_4);j\rangle ^{E} \qquad\;\;   \label{eq:MK}
\eea

The entropic correlation $M(E)$ is strictly positive and this is the only contributing term to the response in equilibrium. However, as the driving field $E$ is increased,  a finite contribution $K(E)$ to the response gets established. We use numerical simulations to get quantitative result for the various correlations in Eq. \eqref{eq:MK}. Fig. \ref{fig:ST}(b) shows plots of $M(E)$ and $K(E)$ as functions of $E.$ The negative frenetic term $K(E)$ overcompensates the entropic component and eventually makes the differential conductivity negative.



\begin{thebibliography}{10}


\bibitem{Green1954} M. S. Green, 
J. Chem. Phys {\bf 22},  398 (1954);
---, J. Chem. Phys. {\bf 19}, 1036 (1951); ---, Phys. Rev. {\bf 119}, 829 (1960).

\bibitem{Kubo1957} R. Kubo, 
J. Phys. Soc. Jpn. {\bf 12}, 570 (1957).

\bibitem{mori}
H.~Mori, Phys. Rev. {\bf 112}, 1829 (1958).

\bibitem{lar} H.~Spohn, {\it Large scale dynamics of interacting particles}.  Springer-Verlag, 1991.

\bibitem{helf} E.~Helfand, Phys. Rev. {\bf 119}, 1 (1960).


\bibitem{Cleuren} B. Cleuren and C. Van den Broeck,  \pre {\bf 67}, 055101 (R) (2003).
 
\bibitem{Reimann} P. Reimann, R. Kawai, C. Van den Broeck and P. H\"anggi, Europhys. Lett., {\bf 45}, 545 (1999).

\bibitem{spiechowicz} J. Spiechowicz,  J. {\L}uczka and P. H{\"a}nggi, J. Stat. Mech. P02044 (2013). 

\bibitem{Zia} R.K.P.~Zia,  E.~L.~Pr{\ae}stgaard, and O.G.~Mouritsen, 
Am. J. Phys. {\bf 70}, 384 (2002). 


\bibitem{fdr} M. Baiesi, C. Maes and B. Wynants, 
Phys. Rev. Lett. {\bf 103}, 010602 (2009).

\bibitem{update} M. Baiesi, C. Maes, New J. Phys. {\bf 15}, 013004 (2013).


\bibitem{time} C.~Maes and K.~Neto\v{c}n\'y, 
J. Stat. Phys. {\bf 110}, 269 (2003).


\bibitem{leb} P.G.~Bergman and J.L.~Lebowitz, 
Phys. Rev. {\bf 99}, 578 (1955).

\bibitem{kls}
S.~Katz, J.L.~Lebowitz, and H.~Spohn, 
J. Stat. Phys. {\bf 34}, 497 (1984).


\bibitem{har} T.~Harada and S.-I.~Sasa, 
Phys. Rev. Lett. {\bf 95}, 130602 (2005).

\bibitem{der} B.~Derrida, 
J. Stat. Mech. P07023 (2007).


\bibitem{hal} H.~Tasaki, 
{\tt arXiv:0706.1032v1}. 


\bibitem{Barma} M. Barma and D. Dhar, J. Phys. C {\bf 16}, 1451 (1983).

\bibitem{zwan}R. Zwanzig, J. Phys. Chem., {\bf 96}, 3926 (1992).


\bibitem{ghosh2012brownian} P. K. Ghosh, P. H{\"a}nggi, F. Marchesoni, F. Nori, and G. Schmid, Phys. Rev. E  {\bf 86}, 021112 (2012).

\bibitem{ghosh2012driven}P. K. Ghosh, P. H{\"a}nggi, F. Marchesoni, S. Martens, F. Nori, L. Schimansky-Geier, and G. Schmid, Phys. Rev. E  {\bf 85}, 011101 (2012).


\bibitem{prs} M. Baiesi, C. Maes and B. Wynants, 
Proc. R. Soc. A {\bf 467}, 2792 (2011).

\bibitem{soghra} C. Maes, S. Safaverdi, P. Visco and F. van Wijland, 
Phys. Rev. E {\bf 87}, 022125 (2013).

\bibitem{Lorentz} H. A. Lorentz, Proc. R. Acad. Sci. Amsterdam {\bf 7}, 438 (1905).


\bibitem{Beijeren} H. van Beijeren, Rev. Mod. Phys. {\bf 54}, 195 (1982). 

\bibitem{Dettmann} C. P. Dettmann  
in {\it Hard ball systems and the Lorentz gas}, edited by D. Sz\'{a}sz, Springer (2000).


\bibitem{Li} B. Li, L. Wang, G. Casati, Appl. Phys. Lett. {\bf 88}, 143501 (2006).

\bibitem{Yang} N. Yang, N. Li, L. Wang, and B. Li, Phys. Rev. B {\bf 76}, 020301(R) (2007).

\bibitem{Hu} B. Hu, D. He, L. Yang, and Y. Zhang, Phys. Rev. E {\bf 74}, 060101(R) (2006).

\bibitem{Shao} Z. G. Shao, L. Yang, H. K. Chan, and B. Hu, Phys. Rev. E {\bf 79}, 061119 (2009).

\bibitem{LiRMP} N. Li, J. Ren, L. Wang, G. Zhang, P. H\"anggi, and B. Li, Rev. Mod. Phys. {\bf 84}, 1045 (2012).  


\bibitem{negcap} E.~Boksenbojm, C.~Maes, K.~Neto\v{c}n\'y and J.~Pe\v{s}ek, 
Europhys. Lett. {\bf 96}, 40001 (2011).

\bibitem{jirka} J.~Pe\v{s}ek, E.~Boksenbojm, and K.~Neto\v{c}n\'y,
Cent. Eur. J. Phys. {\bf 10}, 692 (2012). 


\bibitem{markov} M. Baiesi, C. Maes, B. Wynants, 
 J. Stat. Phys. {\bf 137}, 1094 (2009).








\end{thebibliography}
\end{document}